\documentclass{article}
\usepackage[utf8]{inputenc}
\usepackage{graphicx}
\usepackage[ruled]{algorithm2e}
\usepackage{amsmath}
\usepackage{amssymb}
\usepackage{wrapfig}
\usepackage{float}
\usepackage{hyperref}
\usepackage{booktabs}
\usepackage{threeparttable}
\usepackage{siunitx}
\usepackage{amssymb}
\usepackage{bm}
\usepackage{tcolorbox}
\usepackage[papersize={8.5in,11in},margin=1in]{geometry}
\usepackage{subcaption}
\usepackage{newunicodechar}
\usepackage{array} 
\usepackage{makecell}
\usepackage{microtype}

\usepackage[
backend=biber,
]{biblatex}
\providecommand{\keywords}[1]
{
  \small	
  \textbf{\textit{Keywords---}} #1
}

\usepackage{newunicodechar}
\DeclareRobustCommand{\okina}{%
  \raisebox{\dimexpr\fontcharht\font`A-\height}{%
    \scalebox{0.8}{`}%
  }%
}
\newunicodechar{ʻ}{\okina}

\graphicspath{{figures/}}

\title{The Search for Squawk: Agile Modeling in Bioacoustics}

\author{ \bf
Vincent Dumoulin$^{a\dagger}$,
Otilia Stretcu$^{b\dagger}$,
Jenny Hamer$^a$,
Lauren Harrell$^b$,\\ \bf
Rob Laber$^b$,
Hugo Larochelle$^a$,
Bart van Merriënboer$^a$, \\ \bf
Amanda Navine$^c$,
Patrick Hart$^c$,
Ben Williams$^d$,\\ \bf 
Timothy A.C. Lamont$^e$,
Tries B. Razak$^f$,
Mars Coral Restoration Team$^g$,\\ \bf 
Sheryn Brodie$^i$,
Brendan Doohan$^h$,
Phil Eichinski$^h$,\\ \bf 
Paul Roe$^h$,
Lin Schwarzkopf$^i$,
Tom Denton$^a$ \\ \\
$\dagger$ Indicates shared first authorship.\\
$^a$Google Deepmind,
$^b$Google Research,\\
$^c$Listening Observatory for Hawaiian Ecosystems (LOHE), University of Hawai‘i at Hilo,\\
$^d$University College London,
$^e$Lancaster University UK,
$^f$IPB Indonesia,\\
$^g$Mars Sustainable Solutions, Indonesia,
$^h$Queensland University of Technology,\\
$^i$James Cook University
}

\begin{document}
\maketitle
\begin{abstract}
Passive acoustic monitoring (PAM)  has shown great promise in helping ecologists understand the health of animal populations and ecosystems. However, extracting insights from millions of hours of audio recordings requires the development of specialized recognizers. This is typically a challenging task, necessitating large amounts of training data and machine learning expertise. In this work, we introduce a general, scalable and data-efficient system for developing recognizers for novel bioacoustic problems in under an hour. Our system consists of several key components that tackle problems in previous bioacoustic workflows: 1) highly generalizable acoustic embeddings pre-trained for birdsong classification minimize data hunger; 2) indexed audio search allows the efficient creation of classifier training datasets, and 3) precomputation of embeddings enables an efficient active learning loop, improving classifier quality iteratively with minimal wait time. Ecologists employed our system in three novel case studies:
analyzing coral reef health through unidentified sounds; identifying juvenile bird calls in Hawai‘i to quantify breeding success and improve endangered species monitoring; and Christmas Island bird occupancy modeling.
We augment the case studies with simulated experiments which explore the range of design decisions in a structured way and help establish best practices. Altogether these experiments showcase our system’s scalability, efficiency, and generalizability, enabling scientists to quickly address new bioacoustic challenges.

\end{abstract}

\section{Introduction}


Passive acoustic monitoring (PAM) provides an unparalleled and dynamic view into ecosystems, giving insights into the distribution and behavior of wildlife~\cite{sugai2019terrestrial}. PAM deployments regularly produce thousands---or even millions---of hours of data, requiring machine assistance to uncover these insights.

The extensive diversity of life on earth means that new questions are constantly arising. For many species, especially in the tropics where diversity is greatest, acoustic training data for species classification and detection is hard to come by. Moreover, we are often interested in monitoring specific behaviors, which can be indicated by particular kinds of vocalizations: juvenile bird calls may be used to track year-on-year reproductive success and population health, and territorial calls may be used to identify breeding sites and status~\cite{teixeira2019}. Additionally, geographic variability often makes local classifiers desirable: migratory birds often utilize different vocalizations in breeding and wintering areas, and many species exhibit geographic variability in territorial vocalizations~\cite{wonke2009song}. Finally, in some cases we may not know which sounds originate from which species—for example, fish in coral reefs are difficult and expensive to observe.

This wide range of research questions requires flexible and general machine learning systems, adaptable to new signals with minimal effort~\cite{stowell2022computational}, but also highlights the importance of experts in local ecosystems to select and group meaningful signals for monitoring efforts. The efficient development of bioacoustic classifiers can help address a wide range of problems by providing specific and actionable detections. Outputs from novel classifiers can be used to improve species presence discovery or to understand covariate impacts, ecosystem health, and more. However, gathering extensive training data and managing classifier training remains a challenge for practitioners. Our core goal, then, is to produce a system for bioacoustic analysis which is {\bf efficient} for the end-to-end development of novel classifiers, {\bf adaptable} to new signals and contexts, and {\bf scalable} to very large datasets. Finally, the produced classifiers should have sufficient {\bf quality} for downstream conservation and ecology applications.

\begin{figure}[t]
    \centering
    \includegraphics[width=\textwidth]{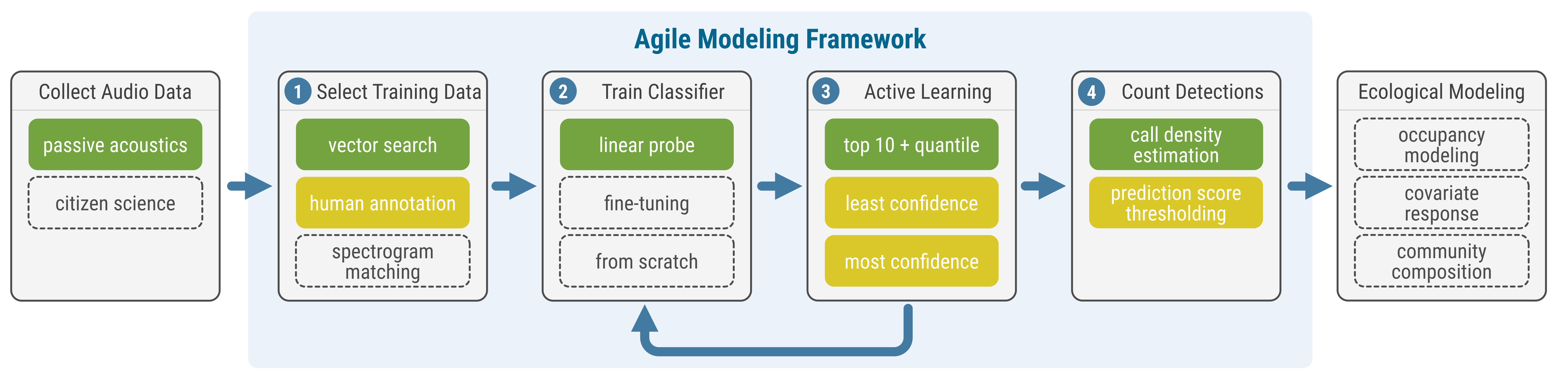}
    \caption{Workflow for the development of novel bioacoustic classifiers. In this paper, we focus on steps 1 through 4. For each step, many strategies are possible, with some options listed below each step. The light-green boxes indicate the strategies we highlight in this paper. We provide explorations and comparisons with the strategies in yellow boxes.}
    \label{fig:workflow_diagram}
\end{figure}

Figure~\ref{fig:workflow_diagram} outlines a general workflow for ecosystem understanding with passive bioacoustic data. Our proposed approach (Agile Modeling) is highlighted in the green boxes. After the collection of audio data, initial annotations are collected (step 1), after which the workflow enters an active learning loop where a new classifier is trained (step 2) on a combination of the annotated examples collected so far and newly human-annotated examples surfaced from the previous classifier's predictions (step 3). Once a classifier with a satisfactory level of performance is obtained, it is used to process the target dataset (step 4) into inputs amenable to downstream ecological analysis, e.g. to evaluate animal occupancy, abundance, reproductive success, community composition, and more. Each step in the development of a classifier (1--4) admits a range of potential strategies, each of which has implications for our key goals of {\bf efficiency}, {\bf adaptability}, {\bf scalability}, and {\bf quality}.

Various forms of human-in-the-loop workflows above have already been used for ecosystem understanding. 
In the broader biodiversity and conservation setting, active learning has been employed to estimate geographic ranges of species~\cite{lange2023activelearning-species}, which is of particular value when observations for a species of interest are scarce. Remote sensing---the use of satellite or aerial imagery---combined with a human-in-the-loop analysis method has allowed for more effective and scalable automated whale monitoring~\cite{boulent2023scaling}.
Active learning has been shown to be useful in processing PAM data using classical machine learning algorithms~\cite{kholghi2018active}, even without the use of neural networks. 
Other workflows have involved template matching to build a CNN-based multi-class, multi-label model to process soundscape data~\cite{lebien2020pipeline}, and using ecoacoustic analysis platforms such as ARBIMON~\cite{aide2013real} to generate labeled data for model training.

Active learning has previously been used to reduce the need for expert human annotation in bird classification~\cite{qian2017active}, and to monitor rare species~\cite{mcewen4767161active,van2023active}.
Search provides efficient generation of initial training data~\cite{allen2025use}, and transfer learning removes the lag between annotation and evaluation of an improved model.
Prior work~\cite{van2023active, allen2021convolutional} leveraged pre-trained embeddings not specific to animal vocalizations and had several manual iterations between reviewed label sets and model updates. 
Recently, there has been initial work combining transfer learning and active learning to process PAM data~\cite{kath2024leveraging}, using academic datasets to validate the technique. We go beyond this work by demonstrating a much broader adaptability of our workflow across distinct applications, providing an extremely efficient process for domain experts to use on their target tasks and at an unparalleled scale as demonstrated by our challenging real-world case studies.

\subsection{Agile modeling}

Our proposed implementation of the workflow, termed {\bf agile modeling}~\cite{stretcu2023agile}, improves on prior active learning workflows through the use of a static bioacoustic foundation model (i.e., fixed embeddings derived from a global birdsong signal classifier rather than a more general audio representation) and the precomputation of embeddings. This dramatically accelerates the workflow, reducing expert review and model update times from weeks to under an hour.

Agile modeling first {\bf precomputes embeddings} from the collected audio data using the foundation model, which crucially only serves as a feature extractor and is {\em not} adapted in subsequent stages of the workflow. This enables an efficient and scalable active learning loop: if the embedding function itself changes as a result of training the classifier (for example, if the classifier takes raw audio as input and uses a composition of the embedding function and a linear classifier for prediction), each active learning cycle would require re-processing the entirety of the collected data---which may take several days when working with large datasets---thereby impacting scalability. By contrast, high-quality static embeddings allow an uninterrupted active learning loop, reducing search and classification retrieval to seconds. Additionally, if the audio data is used for multiple purposes (e.g., if someone later revisits the dataset to detect a different species or call-type), the precomputed embeddings are immediately available, making efficient use of computation through amortization.

Our approach then builds an initial training dataset through {\bf vector search}~\cite{stretcu2023agile,guo2020accelerating}: a user-provided search query (in the form of an example audio clip) is embedded using the bioacoustics foundation model and compared against the precomputed embeddings, and the audio associated with the most similar embeddings is presented to the user for annotation. This serves as an efficient and scalable alternative to brute-force human review: instead of sequentially scanning through the collected audio data in search for multiple positive examples of the vocalization of interest, the user only has to provide a {\em single} positive example and annotate the surfaced audio clips. Such an approach usually faces a chicken-and-egg problem: one needs a good existing classifier to retrieve many good candidates, but one also needs many previously-annotated examples to train a good classifier in the first place. Previous work has shown that global birdsong classifiers provide highly transferable features for novel animal vocalization classification problems~\cite{ghani2023} (and furthermore have the potential to specialize to new contexts~\cite{surfperch}), and this property is what allows us to break out of this conundrum. The broad transferability of these features to novel problems stems from the sheer diversity and complexity of global bird classification, which---as we demonstrate empirically---is also crucial for the overall success of the agile modeling workflow.

A {\bf simple (often linear) classifier} is then trained on the annotated embeddings. Doing so (rather than fine-tuning the embedding function or training a deep neural network classifier from scratch) is extremely efficient: since the embeddings are precomputed and static, the classifier can be trained in usually less than a minute without requiring any special accelerator hardware. The simple classifier is itself not very expressive, but relying on the bioacoustic foundation model to precompute embeddings means that the features on which the classifier is trained are often highly separable for the downstream applications of interest~\cite{ghani2023}.

Finally, the active learning loop is closed using a strategy for presenting embeddings to the user for annotation that uses the classifier outputs to produce a set of candidates that combines top-scoring embeddings and embeddings from a wide range of score quantiles (``{\bf top 10 + quantile}'').

\subsection{Case Studies}

To demonstrate the effectiveness of our system, we present three distinct case studies using real passive acoustic datasets and fixed (precomputed) embeddings. In each case, agile modeling is used to efficiently develop novel, high-quality classifiers, thereby highlighting the system's desiderata and illustrating how novel classifiers can be used to improve ecological understanding.

\paragraph{Hawaiian honeycreepers} The decline of Hawaiian endemic bird species is primarily attributed to loss of habitat and the expansion of invasive mosquitoes. Mosquitoes carry introduced avian diseases such as avian malaria which are often deadly for endemic Hawaiian species~\cite{Warner_1968}. Honeycreeper species have historically been difficult to monitor with bioacoustics approaches because their vocalizations are often confused~\cite{Kahl_2022}. Furthermore, hatching-year Hawaiian honeycreepers suffer the greatest losses during malarial epidemics~\cite{vanRiper_1986}, and as a result juvenile vocalization monitoring could be used to assess the efficacy of mosquito control efforts, with increased juvenile call densities serving as an indicator of reduced disease prevalence~\cite{Gilman_2007}. We develop classifiers with adult and juvenile categories for two distinct honeycreeper species. This demonstrates both the ability of our system to unlock population health and behavioral monitoring and its adaptability to more granular vocalization categories.

\paragraph{Large-scale coral reef restoration} Monitoring represents a significant bottleneck for most reef restoration projects. This means most projects fail to measure their impact, and those that do primarily focus on coral related metrics such as percentage cover and growth which overlook the broader ecological community~\cite{Bostrom-einarsson_2020}. This monitoring bottleneck prevents the dissemination of which techniques work and where. Furthermore, most reef restoration projects are small in scale, but large scale efforts are set to grow in the coming years~\cite{Duarte_2020}. A key challenge is determining whether active reef restoration can recover a fully functioning ecosystem comparable to naturally healthy baselines, and, if so, how to scale these efforts to achieve significant impacts. The soundscape of a coral reef contains information that can both operate as an indicator of reef health---through measuring the abundance and diversity of biophony~\cite{Lamont_2022}---and a function in its own right that mediates recruitment of juvenile fish, corals and other invertebrates~\cite{Simpson_2005,Butler_2022}. Bioacoustics therefore represents a useful medium through which to monitor the broader ecological community and its functioning to determine the success of restoration interventions. We create classifiers for fish sounds in a coral reef environment to track the outcome of large-scale reef restoration and demonstrate not only our system's potential in guiding ecological research, but also its ability to work in a very different environment (underwater) and for sound categories whose biological origin remain to be determined. 


\paragraph{Christmas Island} Monitoring birds on offshore islands is critical for conservation, as many islands support unique and threatened species. Christmas Island is home to seven endemic birds, many of which are threatened or endangered, by habitat loss, invasive species, climate change and disease~\cite{southwell2024power}. Because of the  remoteness  of the  island, these birds are difficult to survey regularly, although monitoring detectability is highly desirable both to discover declines and to quantify positive responses to conservation interventions (e.g.,~\cite{tiernan2025stable}).  In addition, comprehensive bird surveys of the island require at least one highly trained birdwatcher to survey for 5--6 weeks to obtain sufficient data for occupancy modeling~\cite{southwell2024power}.  Inaccessibility has provided motivation for a project trialing passive acoustic monitoring, which has produced hundreds of thousands of hours of recordings.  Further, difficulty accessing the island means there was almost no acoustic data available in public archives for these species. We develop classifiers for three of these low-data species using agile modeling and demonstrate the method's ability to scale occupancy modeling to this large dataset.

\paragraph{} Together, all three studies highlight the generalizable nature of the system, with the coral reef study demonstrating a particularly large domain shift. For all case studies, we collected human validation timing data for the development of the classifiers, demonstrating the efficiency of the system. The Christmas Island study shows scalability to large datasets, and all studies obtained classifiers of sufficient quality for their needs.

\subsection{Simulated Experiments}

We augment the three case studies with simulated experiments which explore the range of design decisions in a structured way and help establish best practices. For these experiments, we work with existing fully annotated bioacoustic datasets to simulate the entire agile modeling approach end-to-end, using the data annotations as proxies for human validation decisions. We work with three datasets: a Hawaiian dataset (which coincides with the Hawaiian case-study)~\cite{hawaiidata2022}, Anuraset (demonstrating the capacity to work with non-bird data)~\cite{canas2023anuraset}, and a Pacific-Northwest bird call-type dataset (demonstrating finer-grained classification, below the species level)~\cite{weldycalltype2024}. The simulated studies explore best-practices for active learning on bioacoustic datasets, which typically have extreme and unpredictable label imbalance~\cite{birdsbatsbeyond}. We also explore the impact of deciding to build a single species-level classifier compared to aggregating outputs from multiple call-type classifiers.

\section{Methods}

This section provides the specifics on the agile modeling workflow for bioacoustics (Steps 1, 2 and 3) and call density estimation (Step 4). We also describe the the simulated experiment protocol, and details on the three human-in-the-loop case-studies, which detail specific ecological modeling applications.

\subsection{Agile Modeling}
\label{ssec:agile_methods}

In the agile modeling workflow, we rapidly develop a classifier for a target sound within a fixed dataset. An embedding model is used to produce feature vectors for each window of audio in the dataset. The process consists of one or more searches for relevant data, then training a classifier, and then iteratively validating data and re-training the classifier in an active learning workflow.

Once the classifier is complete, audio windows containing the target sound are counted, either by selecting a threshold and counting windows with classifier scores above the threshold, or using call-density estimation (Section~\ref{sssec:call_density}).

\textbf{Audio Embeddings.} Any model can be used to produce audio embeddings. For our Hawai`i and Christmas Island case-studies, we use embeddings from the the Perch birdsong classification model. For the Coral Reef case studies, we use the SurfPerch model, which is trained on a combination of bird and reef data. In both cases, the embedding model consumes 5-second audio windows at 32 kHz sample-rate and outputs a 1280-dimensional feature vector for each audio window. The audio from each case study has varying characteristics (described in Section~\ref{ssec:casemethods}), and was resampled to 32 kHz as-needed.

In our simulated data experiments (Section~\ref{ssec:simulated}), we test a wider range of audio embedding models.

\textbf{Vector Search for Training Examples} Training examples for the classifier were produced by brute vector search. An embedding is computed for an initial piece of query audio provided by the practitioner. This query embedding is then compared with each embedding from the target dataset, and ranked according to its inner product.

We find that brute force vector search can process over 1 million embeddings per second (corresponding to around 1,500 hours of audio) on a modern personal computer. Thus, brute-force search only becomes inconvenient when working with over 100,000 hours of audio. The simulated experiments, Hawai‘i, and Coral Reef case studies all work with less than 10,000 hours of audio. The Christmas Island study worked with 10,000 hours of randomly sub-sampled embeddings for classifier development.


\textbf{Classification.} Given a set of annotated embeddings, we train a linear classifier. Previous work has shown that global birdsong embeddings enable few-shot learning on a variety of novel problems, including call-types and vocalizations of non-bird taxa~\cite{ghani2023}, using simple linear models. All-against-all training on a complicated set of problems enables exceedingly simple few-shot learning with linear probes~\cite{chen2019closerfewshot, frustrating2020}.

From the available annotated data, a random train/validation split is produced, with 90\% of the positive data for each class used for training, and the remainder for validation. For the case studies, the classifier is trained with logistic regression. Because examples can contain multiple classes, no softmax is applied. We use mini-batch gradient descent for training the classifier. The trained linear classifier is then tested on the held-out data, and the model ROC-AUC is displayed to the user.

\textbf{Active Learning} Once trained, the classifier is used to surface results to the user for validation. The classifier is run over the entire dataset, and a histogram of the output scores is produced and shown to the user. The user is given the choice of validating the highest-scoring examples or the examples with scores nearest to a user-selected target score. After validating a set of examples, the validated points are included in the training set and a new version of the classifier is trained.

\subsection{Call Density Estimation}
\label{sssec:call_density}

Given a trained classifier, call density estimation approximates the overall prevalence of the target signal in the target dataset while simultaneously producing a model quality estimate~\cite{Navine_2024}. The call-density is the probability that any given audio example in the dataset contains the target signal, which we denote $P(\oplus)$. Briefly, examples are bucketed into logarithmic quantiles (i.e., bottom 50\%, next 25\%, next 12.5\%, etc.) according to the classifier score, and then a reviewer validates $K$ examples from each bucket. We then decompose $P(\oplus)$ over the quantile buckets $b$:
\[
  P(\oplus) = \sum_b P(\oplus|b) P(b).
\]
Here $P(b)$ is known by construction, and $P(\oplus|b)$ is modeled as a beta distribution using the validation results. Bootstrap sampling from the component beta distributions then yields an estimated distribution for $P(\oplus)$, sampling from the component beta distributions 10,000 times.

The same validation data can also be used to estimate the model's area under the receiver operating characteristic curve (ROC-AUC), which admits a probabilistic interpretation as the probability of a positive example having a higher score than a negative example ($P(\oplus > \ominus)$)~\cite{birdsbatsbeyond}. For this, the ROC-AUC is decomposed over quantiles $b$, $c$ as:
\begin{align*}
P(\oplus > \ominus) & = \sum_{b,c} P(\oplus_b > \ominus_c) P(\oplus \in b, \ominus \in c) \\
& = \sum_{b,c} P(\oplus_b > \ominus_c) P(b|\oplus) P(c|\ominus),
\end{align*}
where $\oplus_b$ is shorthand for the set of positive instances in quantile $b$, and $\ominus_c$ is similarly defined. Then $P(\oplus_b > \ominus_c)$ is the probability that the score of a positive example in quantile $b$ is greater than the score of a negative example in quantile $c$. This is identically one or zero when $b \neq c$, and is a ``local'' ROC-AUC when $b=c$. For the ROC-AUC estimate, we use the expected value of $P(\oplus|b)$, $P(\ominus|b)$ and use Bayes rule to obtain $P(b|\oplus)$ and $P(c|\ominus)$.

\subsection{Human Validation Timing Experiments}

To demonstrate the flexibility and speed of our proposed workflow, we conducted timing experiments alongside each of the three case studies. Overall classifier development speed depends on a combination of algorithmic efficiency and human validation speed. To understand the overall efficiency of the process, we separate measurement of machine characteristics (e.g., search latency) from human characteristics (validation speed). The development process is non-linear, and may involve different numbers of searches and active learning refinement steps for different classes, depending on factors such as intra-class variation and problem difficulty.

Each case study engaged in classifier development on a collection of domain-specific signals, logging the steps taken and amount of time spent validating search and classification results, and performing final call-density estimation and model quality assessment. 

The Coral Reef and Christmas Island applications worked with a large combined dataset, iteratively using search and active learning until a satisfactory model was produced, following the procedure outlined in Section~\ref{ssec:agile_methods}.  

By contrast, the Hawai‘i group targeted a fixed number of training examples per classifier, and validated the top 500 search results for a variety of queries until the target number of positive examples had been discovered. These examples were then used to train a classifier, which was then validated using the Call Density Estimation method.

The Hawai‘i group also performed a timing experiment for data collection effort required by linear scan of data: After heuristically identifying a collection of recordings likely to contain the target classes, volunteers listened and labeled each recording. This allows comparison between the amount of time required to surface positive examples with linear scan versus embedding search.

\subsection{Case Study Methods}
\label{ssec:casemethods}

\subsubsection{Honeycreeper Monitoring}
The recordings used in the Hawaiʻi case study were collected at Hakalau Forest National Wildlife Refuge on the eastern slope of Mauna Kea. Hakalau is one of the largest (13,240 ha) intact, disease-free, native forests in the Hawaiian archipelago~\cite{USFWS_2010} and harbors the most intact and stable forest bird community remaining in Hawai‘i~\cite{Kendall_2023}. The refuge is dominated by native ʻōhiʻa (\textit{Metrosideros polymorpha}) and koa (\textit{Acacia koa}) trees, in variable proportion across the landscape, with sparser stands of trees within a pasture matrix at higher elevations, and denser, closed-canopy forests at lower elevations. These recordings were collected year-round at 10 sites along two elevational gradient transects using Songmeter SM4s (Wildlife Acoustics, Maynard, MA USA) as part of a long-term PAM program established by the Listening Observatory for Hawaiian Ecosystems (LOHE Lab) in 2015. In addition to elevation, recording sites along the two transects varied in ecotype and hours of audio data (Table \ref{table:HI_sites}).  

\begin{table}[tbh]
\centering
\begin{threeparttable}
\renewcommand{\arraystretch}{1.3}
\begin{tabular}{@{}llllll@{}}
\toprule
Transect & Site & Elevation (m) & Ecotype                      & Total audio (hours) \\
\midrule
A        & 1    & 1325          & Closed wet ʻōhiʻa forest     & 2127.1 \\
A        & 2    & 1385          & Closed wet ʻōhiʻa forest     & 2121.6 \\
A        & 3    & 1520          & Closed wet ʻōhiʻa forest     & 2197.6 \\
A        & 4    & 1600          & Open wet ʻōhiʻa forest       & 2048.5 \\
A        & 5    & 1818          & Open mesic koa-ʻōhiʻa forest & 2550.7 \\
A        & 6    & 1993          & Open mesic koa pasture       & 2503.9 \\
B        & 1    & 1608          & Closed wet koa-ʻōhiʻa forest & 2019.4 \\
B        & 2    & 1695          & Closed wet koa-ʻōhiʻa forest & 2330.6 \\
B        & 3    & 1874          & Open mesic koa-ʻōhiʻa forest & 3049.8 \\
B        & 4    & 2004          & Open mesic koa pasture       & 1405.1 \\
\bottomrule
\end{tabular}
\caption{\label{table:HI_sites}Hakalau National Forest Wildlife Refuge recording site metadata for the Hawaiʻi case study. Ecotype describes whether the canopy is open or closed, the dominant tree species, and whether the the site is forest or pasture. Total audio describes the hours of recordings collected between 2022--2024.}
\end{threeparttable}
\end{table}

Two experiments were conducted with this dataset and expert annotators to characterize the efficiency and performance of the agile modeling process. In \textbf{Experiment A: Annotation Method Timing}, we compared the review time needed to generate a set of positive examples of an abundant species between traditional approaches of annotating full soundscapes from a linear scan versus results generated from a vector search of a song example from that species. 
In \textbf{Experiment B: Two Species Annotation}, we examined the time to generate a target number of positive examples for adults and juveniles of two endangered species and resulting classifier performance on unseen data for these four classes. 

\textbf{Experiment A: Annotation Method Timing} To compare the vector search method to more traditional manual linear searches of soundscapes, samples were generated for an abundant non-native species, the Red-billed Leiothrix (\textit{Leiothrix lutea}). First, 84 5-min soundscapes (7 hours total audio) from a single recorder collected in February 2024 were opened individually in Raven Pro v1.6~\cite{Cornell_2024} and manually scanned for Red-billed Leiothrix songs, which were selected with annotation boxes with review time recorded. Second, a single vector search was run against a locally sourced Red-billed Leiothrix song example on these same soundscapes, and 500 of the top scoring 5s samples by maximum inner product were reviewed, again logging the human review time. 

\textbf{Experiment B: Two Species Annotation Time~} In this experiment, we focused on the review time, model performance, and downstream ecological analysis of Hawaiian honeycreeper vocalizations. Two species of Hawaiian honeycreeper were included in the Hawaiʻi case study, the ʻAkiapōlāʻau (\textit{Hemignathus wilsoni}), and the ʻAlawī (\textit{Loxops mana}), both of which are federally listed as endangered and considered endangered by the International Union for Conservation of Nature~\cite{IUCN_2024}. 

For this experiment, a custom classifier with two classes (adult and juvenile) was trained for each species---ʻAkiapōlāʻau and ʻAlawī. Vector search was used to collect a target number of training recordings. For the rarer---but more acoustically distinct---ʻAkiapōlāʻau, a training goal was set at 500 samples, while for ʻAlawī, which has similar vocal characteristics to other honeycreeper species, the goal was set at 1,000 samples. Vector search was used to detect target vocalizations within data collected in 2015 and 2016 by two recorders. Setting a large training sample goal and sampling from two recording sites and across multiple years allowed us to capture more acoustic variation in vocalizations.

The preliminary search query examples for each class (one song and one call for adults, and one juvenile begging call) were sourced from Xeno-Canto~\cite{Vellinga_2023} and used in search runs where the top scoring 500 samples, based on maximum inner product, were reviewed from the two recording sites and two years (i.e., search run 1 searched the site 1-2015 data, run 2 searched site 2-2015, run 3 searched site 1-2016, run 4 searched site 2-2016). If after these initial query search runs the training sample goal had not been met, a locally sourced sample derived from the preliminary search was then used as the query example, again reviewing 500 of the top scoring samples from each site/year combination until the target number of samples were achieved for each class. Review time was recorded for each search run. 

Once the target number of honeycreeper training samples were generated using the vector search, we trained  custom linear classifiers on the annotated embeddings with adult and juvenile classes for the two honeycreeper species. We ran each species' classifier over unseen data collected between 2022 and 2024 at a site known to harbor both species and for each target class evaluated the 100 samples with the highest logit scores (logarithm of the odds of the probability of the target species vocalizing). 

The classifiers were then used to estimate call density $P(\oplus)$ for each class at the 10 recording sites along the two transects following the methods described in~\ref{sssec:call_density}. Adult honeycreeper singing rates are highest during the pre-breeding season, as individuals attempt to attract mates and establish or maintain territorial boundaries, then singing drops off precipitously during incubation and nestling care~\cite{Ralph_Fancy_1994}. Juvenile ʻAkiapōlāʻau and ʻAlawī that have fledged from their nests give frequent, distinctive begging calls that serve as contact calls or food solicitation to their parents~\cite{Pratt_2020, Lepson_2020}. For each class, only data collected during months of peak vocal activity were included in call density analysis (December--June for ʻAkiapōlāʻau adults~\cite{Pratt_2020}, June--December for ʻAkiapōlāʻau juveniles~\cite{Ralph_Fancy_aki}, February--June for ʻAlawī adults~\cite{Ralph_Fancy_1994}, May--August for ʻAlawī juveniles~\cite{VanderWerf_1998}). To assess the effect of elevation on call densities we ran a linear model in R (version 4.4.2; \cite{R_2024}) for each class. Data visualization was also conducted in R.

\subsubsection{Coral Reef Health}

Passive Acoustic Monitoring (PAM) data was collected between August 29th and September 23rd at the Mars Coral Reef Restoration Project,\footnote{\url{www.buildingcoral.com}} located west of Bontosua Island, in the Spermonde Archipelago, South Sulawesi, Indonesia ($4^{\circ}55'53.0''\mathrm{S},\; 119^{\circ}18'58.0''\mathrm{E}$). Six naturally healthy (H1--6), six degraded (D1--6), and six restored sites (R1--6) were selected, with a mean coral cover proportion and standard deviation of 0.72\,$\pm$\,0.10, 0.10\,$\pm$\,0.08 and 0.66\,$\pm$\,0.14 respectively. Sites and coral cover measures replicate the 17 study sites used in~\cite{Vida_2024} during the same period, with one additional healthy site to complete the set of six at ($4^{\circ}56'10.1''\mathrm{S},\; 119^{\circ}19'03.8''\mathrm{E}$). Each site consisted of at least a $20\times50$m area of contiguous reef habitat between 2--4 m depth, with a minimum of 20 m distance between adjacent sites. Degraded sites were subject to bomb fishing and coral mining in the 30--40 years prior to the study leading to ecological collapse into rubble dominated reefs~\cite{Williams_2019}. Restored sites were selected from 3.1 ha of previously degraded habitat where the Mars Assisted Reef Restoration technique~\cite{Smith_2021} had been implemented at least three years prior to PAM data collection. 

HydroMoth audio recorders~\cite{Lamont_2022} were placed at the centre of each site and left to record on a one-minute-on, one-minute-off duty cycle for the full 24-day period, with a battery change after seven days. Sites H4 and H5 were the only exceptions, where only six days of audio were collected due to recorder theft after the battery change. Recordings were captured at 16 kHz with ``low-gain mode'' enabled and the lowest of the five gain settings. The final data set consisted of 267,814 one-minute files, totalling 6.11 months of recordings.

\begin{figure}[t]
	\centering
	\includegraphics[width=0.9918\textwidth]{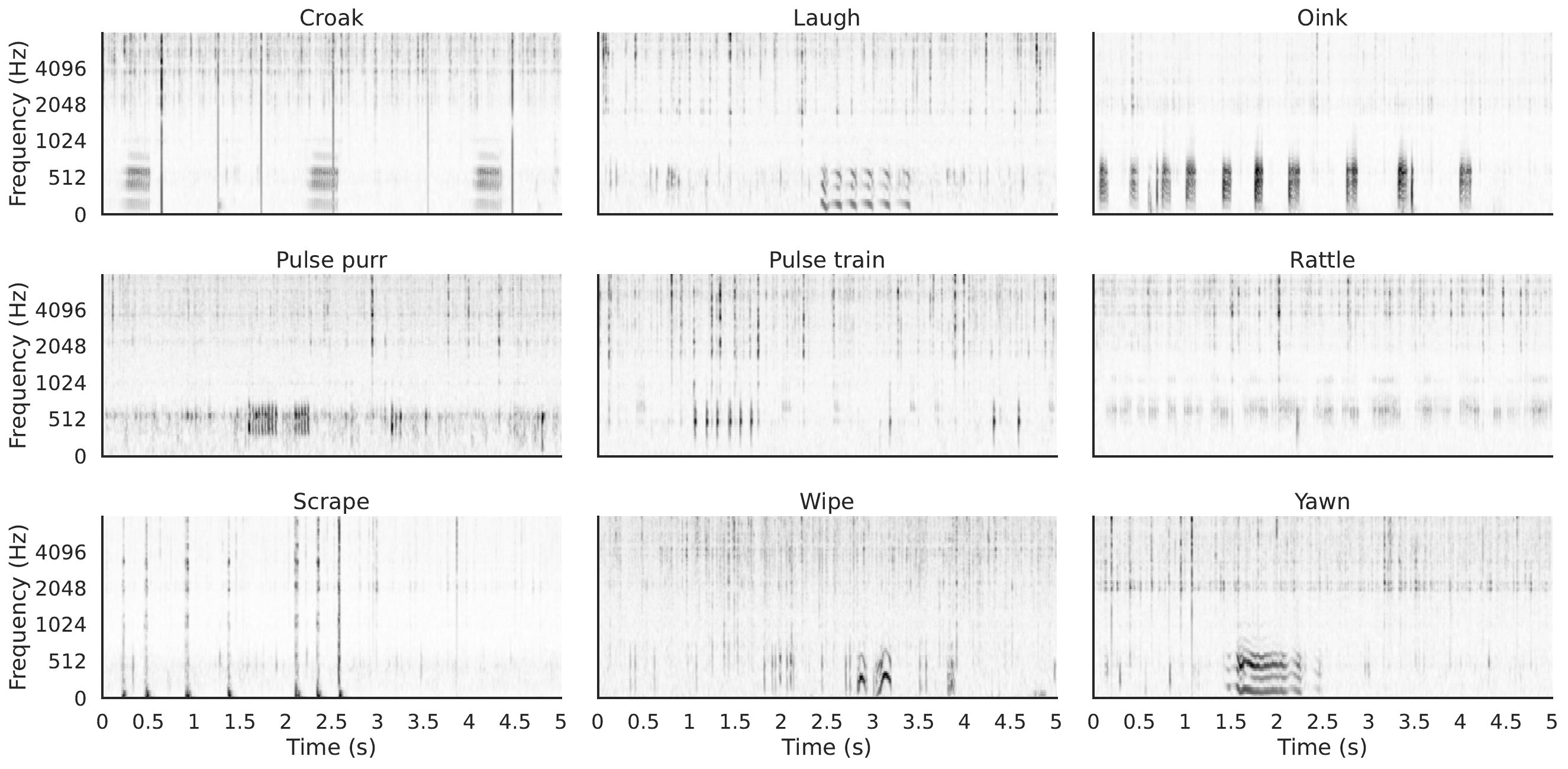}
	\caption{Log-mel spectrogram plots of the nine fish sonotypes identified from the coral reef data set for use as queries.}
	\label{fig:fish_queries_spectrogram_plot}
\end{figure}

Putative fish sonotypes within the data set were manually identified and used as target queries. For the manual identification, 1000 files were generated at random from the data set, balanced by total recording time from each site. The spectrogram of each file was then screened by an expert annotator (BW) to identify fish sounds. Following methods from~\cite{Lamont_2021} and~\cite{Raick_2023}, manually identified fish sounds were reviewed and grouped by their acoustic characteristics into broad sonotype classes using a conservative approach where sounds were only included if there was a very high degree of confidence that they were characteristically different from other sounds. This approach yielded nine sonotypes to be used as queries (\autoref{tab:fish_sounds}, \autoref{fig:fish_queries_spectrogram_plot}). Following query selection, agile modeling was employed following section 2.1 for each sonotype. Here, embeddings were extracted using SurfPerch, a pretrained model optimised for coral reef audio~\cite{surfperch}. 

Given the set of recordings and embeddings, the agile modeling process was then followed, producing classifiers for each sonotype, as well as call-density estimates and classifier quality metrics.

Using the classifiers, we then examine the relationship between different sonotypes and coral reef habitat status, to demonstrate the application to ecological questions. For each site, classifier detections (i.e., audio windows with $>0.5$ classifier confidence) were counted per day. A zero-inflated Poisson distribution was then fit for each site's daily detection counts using an expectation-maximization algorithm. Under this model, each site is parameterized by latent variables $\pi$ (latent presence of the vocalizing species) and $\lambda$ (vocalization rate).

\begin{table}[t]
\centering
\begin{threeparttable}
\renewcommand{\arraystretch}{1.3}
\begin{tabular}{@{}l>{\raggedright\arraybackslash}m{2.5cm}ll>{\raggedright\arraybackslash}m{6cm}@{}}
\toprule
{Name}      & {Frequency range (Hz)} & {Duration (s)} & {Type}       & {Description}                       \\
\midrule
Croak       & 50--850                & 0.4            & Downsweep    & A short downsweeping sound with
                                                                       typically over a dozen harmonics. \\
Laugh       & 100--600               & 1.1            & Downsweep    & Short downsweeping complex tonal
                                                                       sound, repeated three or more
                                                                       times. \\ 
Oink        & 50--1000               & 0.2            & Pulse series & Short rapid pulses, often
                                                                       repeated. \\
Pulse purr  & 300--600               & 0.4            & Pulse series & Short rapid pulses in a series of
                                                                       three or more, often series are
                                                                       repeated. \\
Pulse train & 150--750               & Variable       & Pulse series & Short rapid pulses, three or more.
                                                                       Broader frequency range compared to
                                                                       pulse purr. \\
Rattle      & 450--850               & 0.1            & Pulse series & A repetitive short pulse, often
                                                                       exhibits ``chorusing'' which dominates
                                                                       the soundscape. \\
Scrape      & Broadband              & 0.1            & Percussive   & Crunching sound, as when excavating
                                                                       grazers take bites on hard
                                                                       substrate. \\
Wipe        & 100--600               & 0.3            & Tonal        & Short arched tonal sound, often with
                                                                       harmonic components and repetition. \\
Yawn        & 100--700               & 1.0            & Tonal        & A complex tonal sound with an early
                                                                       rise and decay followed by a longer
                                                                       period of harmonics that decay
                                                                       further at the end. \\
\bottomrule
\end{tabular}
\caption{\label{tab:fish_sounds}Characteristics of the nine putative fish sonotypes identified from the coral reef data set for use as queries, including their frequency range, duration, type, and a brief description.}
\end{threeparttable}
\end{table}

\subsubsection{Christmas Island Bird Monitoring}

Christmas Island is located 200 km south of Indonesia in the Indian Ocean. Its tropical rainforest is home to  a  range  of  endemic  bird  species,  many  of  which  are  threatened.   
Due to the immense time and effort of these manual point-count surveys, acoustic monitoring with automated species call detection is being explored.  Acoustic sensors were deployed at 82 sites between May 2023 and July 2024, recording a total of 268,576 hours of audio. This was produced in two deployments: Deployment 1,  mainly continuous recording from May 2023 to October 2023 (roughly 190,000 hours), Deployment 2, from October 2023 to July 2024, one 20 min recording per hour (roughly 78,000 hours).

Building call-recognizers generally requires extensive labelled example audio clips of the target call-types, however for the Christmas Island recognizers, almost no existing training data was available for three of the species. No example recordings of the Christmas Island Goshawk were available, but examples of the similar-sounding Brown Goshawk were obtained from Xeno-canto. For the Christmas Island Thrush, three examples could be found by laboriously listening through the audio unassisted by automated methods.  For the Christmas Island Emerald Dove, some examples surfaced through verification of false-positive detections of the Christmas Island Imperial Pigeon, which shares some similarities for one of its call types. 

This limited set of recordings was then used as a starting point to build up the labelled training set following the sampled brute-force search process described in Section~\ref{ssec:agile_methods}. After embedding the unlabelled audio, embeddings of the labelled examples were used as search queries, using Maximum Inner Product to compare them with random subsets of the unlabelled set, generally between 5000 and 1000 hours worth, all from Deployment 1. Starting with the highest scoring, these were inspected and labelled positive or negative until at least 20 examples were found. In the case of the Christmas Island Thrush, two different call-types were identified (``chuck'' and ``chitter''). The two call types were queried separately, and labeled as a single class (i.e., using the ``Balanced Search Query'' method explored in the Simulated Experiments). 

For each species, this model was then used to classify examples from a random 10,000 hour subset of the unlabelled set from Deployment 1, following the process described in Section~\ref{ssec:agile_methods}. Examples for active learning were chosen according to the ``least confidence'' criteria, with classifier scores nearest to 0.
This process was repeated between 4 and 6 times with a new random subset of data from Deployment 1 in each round.

\begin{figure}[t]
	\centering
	\includegraphics[width=\textwidth]{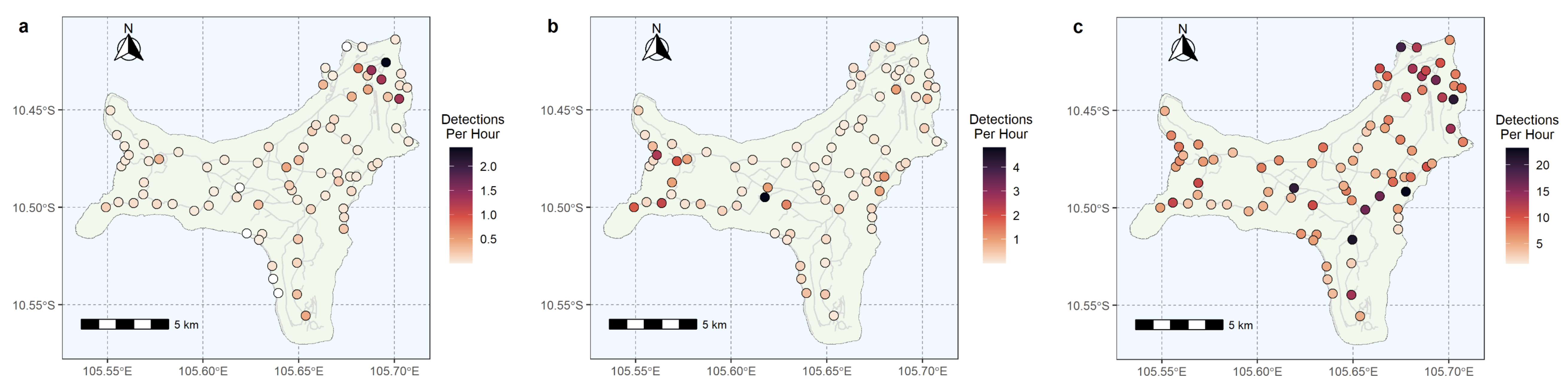}
	\caption{Detection rates across Christmas Island sites for the three species targeted in the study: a) Emerald Dove, b) Christmas Island Goshawk, and c) Christmas Island Thrush.}
	\label{fig:CI_Map_Detection}
\end{figure}

Detection rates were calculated for each site for each of the target species over the whole monitoring period (Figure~\ref{fig:CI_Map_Detection}). High-confidence detections were selected based on analysis of classifier scores on a sample of validated detections, and detection rate was calculated as total number of detections divided by the number of hours of audio recorded and analyzed for that site.

\subsection{Simulated Experiments}
\label{ssec:simulated}

We design simulated experiments to answer the following questions:

\begin{enumerate}
    \item Which active learning strategy should be adopted? Does the optimal choice vary depending on the data regime?
    \item How resilient is the agile modeling workflow to the choice of embedding function? Do representations trained on a global bird song classification task also outperform more general audio representations like prior work~\cite{ghani2023} on transferring to new bioacoustics classification problems suggests?
    \item How should call types be handled when building species-level detectors? Can we successfully combine call type-level detectors into a species-level detector?
\end{enumerate}

Since answering these questions through experiments with human experts would be prohibitively time-consuming, we turn to simulation and propose a methodology that synthesizes human-in-the-loop interactions. Rather than starting with an unannotated soundscape dataset and having a human expert iteratively annotate audio windows surfaced by the workflow, we start with a densely-annotated soundscape dataset whose annotations are initially hidden from the algorithm and simulate annotating a surfaced audio window by revealing its ground truth label. This allows us to measure the performance of the resulting model (in terms of ROC-AUC) throughout the agile modeling workflow and efficiently compare various design decisions. The procedure is summarized as follows:

\begin{enumerate}
    \item Start with a species of interest and a set $\mathcal{D}$ of audio embeddings and their ground-truth annotations (positive or negative) for the species of interest.
    \item Maintain two sets of embeddings: a set $\mathcal{R}$ of embeddings whose labels have been revealed to the model (initially empty), and a set $\mathcal{H}$ of embeddings whose labels remain hidden from the model (initialized with $\mathcal{D}$).
    \item Select one true positive embedding in $\mathcal{H}$ at random to act as the search query and move it from $\mathcal{H}$ to $\mathcal{R}$.
    Score all remaining embeddings in $\mathcal{H}$ according to their inner-product with the search query embedding.
    \item Repeat:
    \begin{enumerate}
        \item Use the scores to select a subset of $\mathcal{H}$ (via some active learning strategy) to ``annotate'' and move to $\mathcal{R}$, thereby simulating their labeling by an expert annotator.
        \item Train a linear classifier on $\mathcal{R}$'s embeddings and ground truth labels.
        \item Score and rank the embeddings in $\mathcal{H}$ using the linear classifier.
        \item Measure the ROC-AUC across $\mathcal{R} \cup \mathcal{H}$ by ranking the embeddings in $\mathcal{H}$ according to their score and placing them in a list between $\mathcal{R}$'s positives (ranked at the very top of the list) and its negatives (ranked at the very bottom of the list).
    \end{enumerate}
\end{enumerate}

Note the decision to measure ROC-AUC across $\mathcal{R} \cup \mathcal{H}$. In this problem, we seek to train a ``domain-specific'' classifier for this particular dataset, surfacing all relevant sounds in-context. Thus, instead of splitting the data into train and test splits, we measure the ultimate retrieval quality of the classifier over the entire dataset. 

\subsubsection{Active learning strategies and embedding quality}

We compare the agile modeling performance of ten embedding functions:
\begin{itemize}
    \item {\bf BEANS baseline}: a hand-designed set of features, which provides a baseline for comparison with learned methods. The features are MFCC statistics (top 20 MFCCs aggregated across time with mean, standard deviation, min, and max operations, for a total of 80 features) used by the BEANS benchmark~\cite{hagiwara2023beans}.
    \item {\bf AudioMAE}~\cite{audiomae} is a self-supervised general audio model with a transformer architecture, trained to predict masked spectrograms. We evaluate `large' and `huge' variants, re-implemented and trained by by Eduardo Fonseca~\cite{georgescu2022audiovisual}. Further information and evaluation of this model in a biouacoustics setting is available in~\cite{ghani2023}.
    \item {\bf BirdNet}~\cite{kahl2021birdnet} is a convolutional bird species classifier, used in a variety of conservation monitoring projects, trained on a combination of open and closed data. We compare versions 2.1, 2.2, and 2.3. Note that the main difference between v2.2 and 2.3 is the embedding dimensionality, which increased from 320 to 1024.
    \item {\bf Perch}~\cite{perch} is, like BirdNET, a convolutional bird species classifier, trained entirely on Xeno-canto. Perch training also includes hierarchical classification of bird genus, family, and order alongside species.
    \item {\bf SurfPerch}~\cite{surfperch}, which is a variant of Perch trained on a combination of bird and reef data.
    \item {\bf YAMNet}~\cite{yamnet} and {\bf VGGIsh}~\cite{vggishmodel} are convolutional models trained on AudioSet. YAMNet uses a MobileNetV1 architecture, while VGGish uses a variant of the VGG architecture and is trained on an early version of AudioSet. Both models process 0.96s frames of audio, and the model embedding dimensions are 1024 and 128 respectively.
    
\end{itemize}
The embedding functions are evaluated across three different densely-annotated soundscape datasets.
\begin{itemize}
    \item {\bf Hawai‘i}~\cite{hawaii_data} is the fully-annotated competition dataset from BirdCLEF 2023. The data includes all species considered in the Hawai‘i case study, but annotated only at the species level.
    \item {\bf Anuraset}~\cite{anuraset} is a fully-annotated dataset containing Colombian frogs (anurans), which we use to demonstrate generalization to non-bird targets.
    \item The {\bf Weldy Call-Type}~\cite{weldy_calltype} dataset consists of 11.75 hours of fully annotated recordings for over sixty species, including annotations for multiple call-types per species. Recordings are from `dawn-chorus' and thus quite dense with annotations.
\end{itemize}

We experiment with four active learning strategies to select a subset of {\bf 50} embeddings in $\mathcal{H}$ to annotate at each round. At a high level, different strategies may be more relevant depending on the classifier quality and difficulty of the concept.

In some cases, there may be negative examples with high scores, in which case it is helpful to surface examples with high activations which will be disambiguated by the human validator ({\bf most confidence}). When the classifier is working well, high-activation examples provide little further information. In this case, validating low-confidence examples ({\bf least confidence}, making use of \textit{margin sampling}~\cite{balcan2007margin}) is a well-known strategy for improving model quality. Conversely, for extremely rare signals, least confidence sampling can struggle to find any further positive examples.

Since we don't know which regime a given classifier may fall into, we also explore two hybrid strategies: logarithmic {\bf quantile} sampling to validate across a range of activations (similar to the validation scheme in~\cite{Navine_2024}), and {\bf top 10+quantile} combining the top-10 results with quantile sampling.


\subsubsection{Call types}

We compare three different strategies to handle call type information by running additional agile modeling simulations on species in the Weldy dataset that have exactly two annotated call types. The strategies we consider are:
\begin{enumerate}
    \item {\bf Balanced search query}: Ensure that the search query contains exactly one vocalization of each call type (two positive embeddings in total), then proceed with the rest of the agile modeling loop while annotating embeddings at the species level (obtaining 50 annotations every iteration).
    \item {\bf Combine call type-level classifiers}: Run an agile modeling loop for each of the two call types in parallel. Each loop is provided with its own search query containing exactly one positive embedding, and to control for the number of annotations, 25 annotations are obtained for each call type at each iteration.
    \item {\bf Control}: Ignore call types altogether and annotate embeddings at the species level. To control for the size of the search query, two randomly-selected positive embeddings for each species are provided (ignoring call type). 50 annotation are acquired at each iteration.
\end{enumerate}

To rank embeddings given multiple scores for each embedding (one for each of the two search query embeddings, or one for each of the call type-level classifier scores), we use a naive heuristic that assigns to each example the largest of its scores (largest inner-product across search query embeddings, or largest logit across call type-level classifiers).

\section{Results}

\subsection{Honeycreeper Monitoring}

In Experiment A: Annotation Method Timing, the manual scan of 84 5-min soundscapes for Red-billed Leiothrix songs in Raven Pro yielded 137 positive samples and took over 4 hours of review (averaging 109.0 seconds of review per positive sample), while review of the top 500 samples from the vector search within these soundscapes took less than 20 minutes and yielded 472 positive detections (averaging 2.5 seconds per positive sample). In other words, the vector search approach was {\bf 43 times faster} than manual review for this abundant species. 

In Experiment B: Two Species Annotation, it took multiple vector search runs---each displaying 500 samples for review---to achieve the training dataset goals for the rarer ʻAkiapōlāʻau and ʻAlawī species (500 and 1000 samples, respectively), with a mean review time of 3.2 seconds per sample reviewed across all classes (Table \ref{table:timings}). 

With the custom species-level classifiers trained for ʻAkiapōlāʻau and ʻAlawī, we ran each over unseen data. All of the 100 top scoring samples evaluated were positive detections for both the adult and juvenile ʻAlawī classes, and the ʻAkiapōlāʻau adult class (precision = 1), while 97 were positive detections of ʻAkiapōlāʻau juveniles (precision = 0.97). Classifier ROC-AUC scores derived from validation results were also high, achieving $\geq 0.81$ for all classes (Table \ref{table:timings}). 

\begin{figure}[thb]
    \centering
    \includegraphics[width=\linewidth]{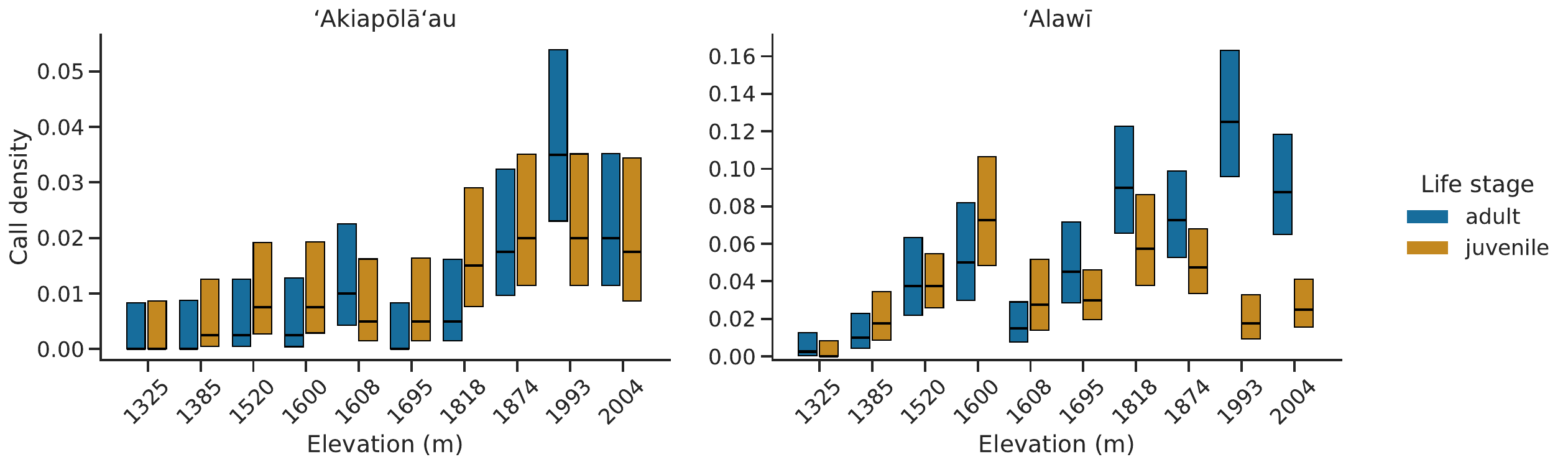}
\caption{Call density estimates across an elevation gradient for two Hawaiian bird species (left ʻAkiapōlāʻau, right ʻAlawī) at two life stages. Blue bars represent adult call densities, and orange bars represent juvenile call densities. Lines through each box represent the maximum likelihood estimate of call density.}
\label{fig:HI_birds}
\end{figure}

We found that elevation was correlated with call density for all classes ($p \leq 0.005$) except ʻAlawī juveniles ($p = 0.481$). ʻAkiapōlāʻau call densities increased with elevation, with neither adults nor juveniles detected at the lowest elevation site (1325 m; Figure \ref{fig:HI_birds}). Juvenile ʻAkiapōlāʻau call densities showed similar patterns to those of the adults, although the high-elevation site with the highest adult call density did not have higher juvenile call densities than the other high-elevation sites. ʻAlawī were detected at all elevations, although only adults were detected at the lowest elevation site. ʻAlawī adult call densities also increased with elevation, but there was more divergence between adult and juvenile call density patterns than observed with ʻAkiapōlāʻau.  


\subsection{Coral Reef Health}
High confidence detections, where the output logit score was $\geq1.0$, totaled to 99,137 across all sites and sonotypes. The mean count of detection's per day was highest on the healthy sites, followed by the restored then degraded sites, with mean and standard deviations of (483 $\pm$ 342), 369 ($\pm$ 143) and (72 $\pm$ 67) respectively. These appeared to be primarily driven by increases in the count of two fish sonotypes on healthy and restored reefs, ‘Pulse train’ and ‘Rattle’. The mean diversity of sonotypes detected each day on restored and healthy reefs was also higher than the degraded reefs, with a mean and standard deviation of 5.13 ($\pm$ 1.27), 5.0 ($\pm$ 1.12) and (3.93 $\pm$ 1.49) respectively.

\begin{figure}[htb]
\centering
\includegraphics[width=0.9987\textwidth]{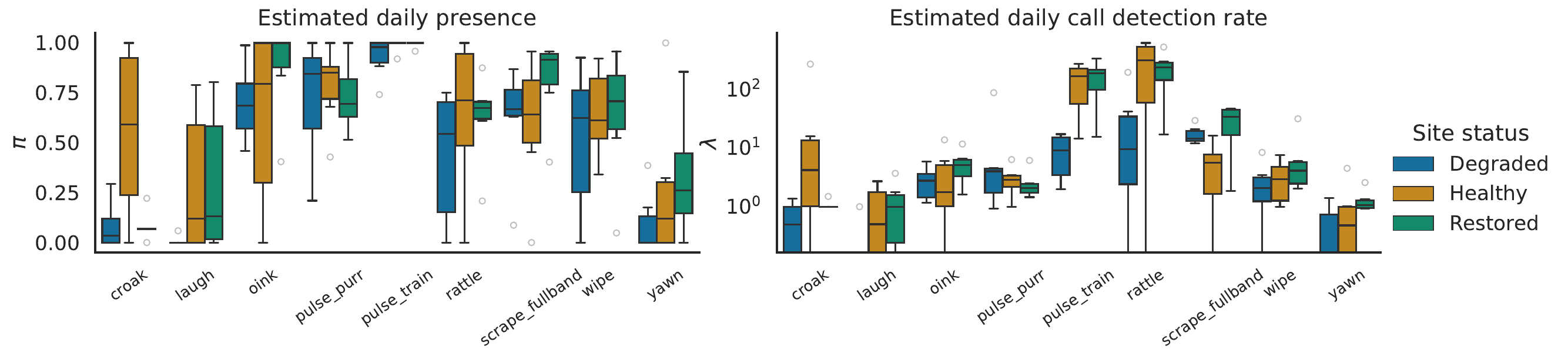}
\caption{Estimated Zero-Inflated Poisson parameters for healthy, degraded, and restored coral reef sites, indicating the latent presence/absence and daily call rates of the target animals.}
\label{fig:reef_zip}
\end{figure}

By fitting the zero-inflated Poisson distributions to each site for each sonotype, we may also examine whether particular sonotype presence absence or detection rates correlate with reef status. The estimated parameters, organized by site status, are illustrated in Figure~\ref{fig:reef_zip}. We notice that detection rates for pulse train and rattle are strong indicators of site health. Interestingly, the `croak' sonotype is more often present and has higher rates at healthy sites than degraded or restored sites.

\subsection{Christmas Island Bird Monitoring}

Call-density validation revealed very high classifier quality, with the worst performance on the Christmas Island Thrush (0.955). There was no noticeable difference in accuracy on Deployment 2. 

Due to the large scale of audio analyzed, the embeddings were generated using high-performance computing (HPC) infrastructure, utilizing the PBS job scheduling system for distributed computation, allowing it to be completed in less than one day. Once embeddings were generated, the time needed to run inference using the precomputed embeddings was very fast, around one hour on a desktop computer.

The Christmas Island Emerald Dove was detected at 77 of the 82 sites (Figure~\ref{fig:CI_Map_Detection}a) whereas the Goshawk (Figure~\ref{fig:CI_Map_Detection}b) and Thrush (Figure~\ref{fig:CI_Map_Detection}c) were detected at all 82 sites throughout the monitoring period. Despite these species being widespread across the island, detection rates varied considerably among sites, revealing variability in site usage. The Emerald Dove, in particular, had low detection rates and was mostly detected at sites in the north-east of the island around the town settlement.

\subsection{Agile Modeling Experiments}

Results of human timings and classifier quality are collected in Table~\ref{table:timings}. We find that human review of examples varied between one and ten seconds per five-second example, averaged over all validation environments (search, classifier outputs, final validation), with an overall average of 4.79 seconds per clip.

For the coral reef and Christmas Island classifiers, multiple searches were sometimes used for a single class to obtain a wider variation of training examples. In both cases, the estimated ROC-AUC from all of the resulting classifiers were very high ($>0.95$). Classifier development when using the iterative process obtained high-quality classifiers in much less than an hour of expert review time per classifier.

By contrast, the Hawai‘i effort used search to obtain a fixed-size training set, and trained a single classifier. The resulting classifiers had very high quality for the ʻAkiapōlāʻau ($> 0.95$), and fairly high quality for the ʻAlawī ($> 0.08$). 

\begin{table}
\centering
\begin{threeparttable}
\renewcommand{\arraystretch}{1.3}
\begin{tabular}{@{}cllllll@{}}
\toprule
Study & Class          & \#S   & \#C   & Total (h) & Per-Example (s)   & ROC-AUC   \\
\midrule
CR & Rattle         & 1     & 6     & 0.46      & 5.11              & 1.000 \\
CR & Oink           & 2     & 5     & 0.69      & 8.27              & 0.994 \\
CR & Croak          & 3     & 3     & 0.34      & 3.49              & 0.980 \\
CR & Scrape         & 2     & 3     & 0.42      & 5.04              & 0.994 \\
CR & Pulse Train    & 2     & 5     & 0.41      & 3.71              & 0.997 \\
CR & Pulse Purr     & 2     & 5     & 0.42      & 3.80              & 0.968 \\
CR & Yawn           & 2     & 4     & 0.35      & 2.83              & 0.988 \\ 
\midrule
CI & Emerald Dove& 1     & 5     & 0.43      & 6.64              & 0.999 \\
CI & CI Goshawk     & 1     & 4     & 0.41      & 7.13              & 1.000 \\ 
CI & CI Thrush      & 3     & 6     & 0.71      & 10.39             & 0.955 \\ 
\midrule
HI & ʻAkiapōlāʻau (Adult)    & 9     & 1     & 3.65*     & 2.92              & 0.982 \\ 
HI & ʻAkiapōlāʻau (Juve)     & 2     & 1     & 1.63*     & 5.87              & 0.967 \\ 
HI & ʻAlawī (Adult)  & 8     & 1     & 2.22*     & 1.99              & 0.826 \\ 
HI & ʻAlawī (Juve)   & 15    & 1     & 4.22*     & 2.03              & 0.886 \\ 
\bottomrule
\end{tabular}
\caption{\label{table:timings}Timing and classifier quality across coral reefs (CR), Christmas Island (CI), and Hawaiian (HI) projects. The total number of searches (\#S), classifier training iterations (\#C), total human review time for search, classification and validation, time spent per-example in review, and the ROC-AUC estimated from validation on field data. *Note that the Hawaiian effort reviewed search results to obtain a fixed number of positive examples, instead of applying an active learning strategy.}
\end{threeparttable}
\end{table}

\subsection{Simulated Experiments}

\begin{figure}[t]
	\centering
	\includegraphics[width=\textwidth]{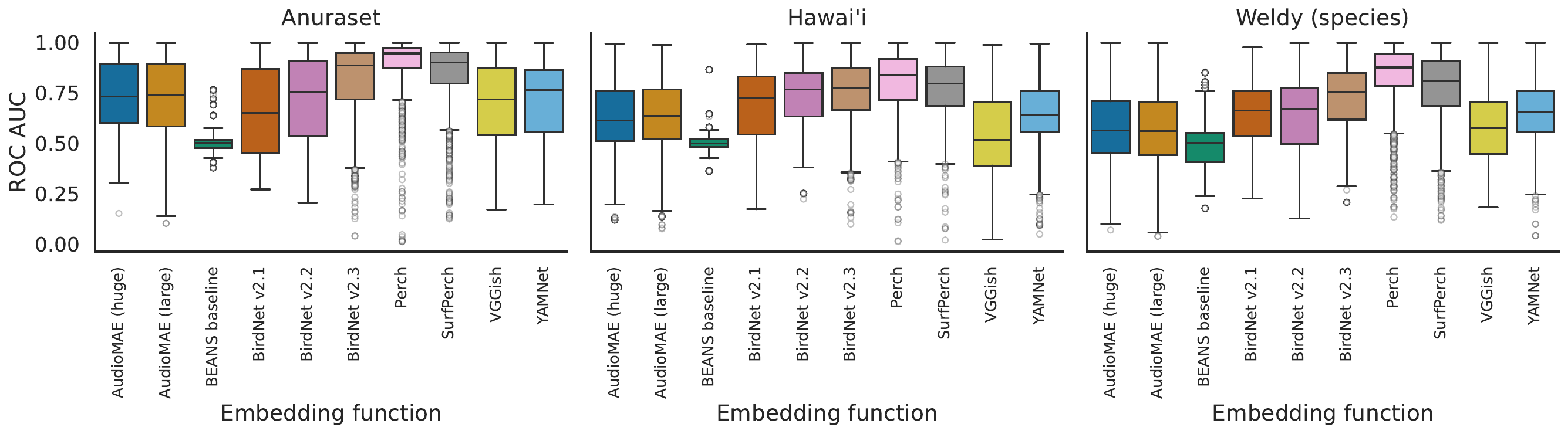}
	\caption{Simulated agile modeling performance across embedding functions and evaluation datasets after 100 annotations have been acquired.}
	\label{fig:embedding_function_comparison}
\end{figure}

We carry out agile modeling simulations for each of the ten embedding functions, each of the four active learning strategies, and for each species in each of the three datasets. We repeat each simulation ten times with a different random seed each time so as to randomize the search query and measure sensitivity to it. This results in 1200 individual simulations, each of which is parallelized over all of its corresponding dataset's species. In each case, we compute ROC-AUC metrics after 100 annotations have been acquired.

\subsubsection{Embedding functions: bioacoustics-specific representations outperform more general representations}

We start by comparing embedding functions across datasets (\autoref{fig:embedding_function_comparison}). On one hand, we observe significant differences in performance across embedding functions, indicating that agile modeling for bioacoustics---unsurprisingly, and like transfer learning~\cite{ghani2023}---is indeed sensitive to the choice of embedding function. On the other hand, the main source of performance variation across embedding functions appears to be the nature of the data that the model was trained on rather than the specific training recipe that was used. Handcrafted features (BEANS baseline) perform very poorly, and embedding functions trained on YouTube-derived datasets (AudioMAE, VGGish, YAMNet), while performing better, do significantly worse than recent embedding functions trained on bird vocalization data (BirdNet v2.3, Perch, SurfPerch).

\begin{figure}[t]
	\centering
	\includegraphics[width=0.6631\textwidth]{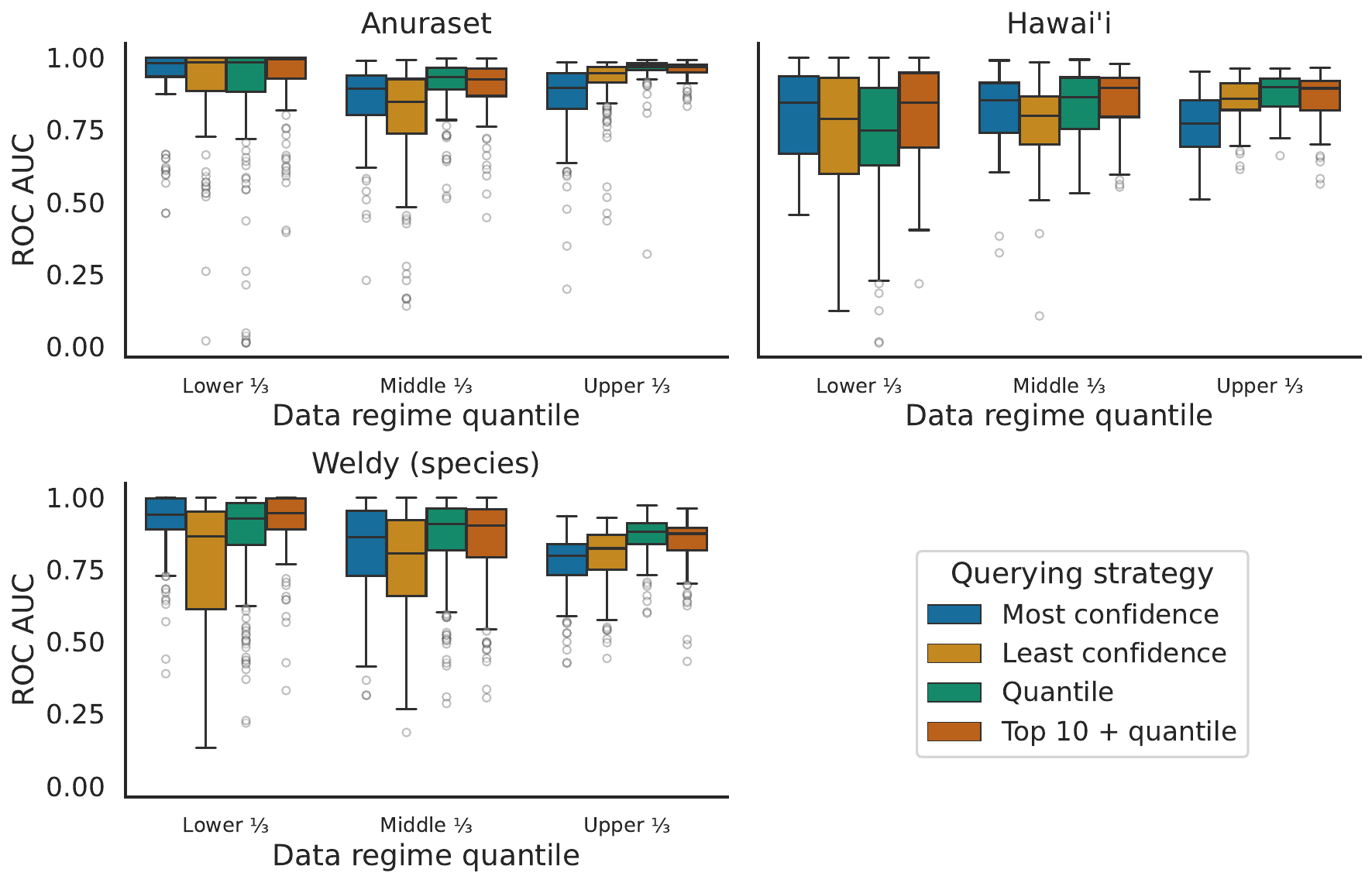}
	\caption{Simulated agile modeling performance for Perch embeddings across data regimes, active learning strategies, and evaluation datasets after 100 annotations have been acquired. The data regimes are represented by three quantiles: lower third, middle third, upper third.}
	\label{fig:querying_strategies}
\end{figure}

\subsubsection{Active learning strategies: ``top 10 + quantile'' strikes a good balance across data regimes}

Using the best-performing embedding function (Perch), we then compare active learning strategies across datasets and data regimes (\autoref{fig:querying_strategies}). For each dataset, we partition species into three bins as a function of each species' true positive rate in the dataset: the lower, middle, and upper third quantiles of true positive rates. We observe that in the medium-to-high data regimes the ``quantile'' active learning strategy performs consistently better than the most and least confidence strategies, whereas in low data regimes the ``most confidence'' active learning strategy frequently outperforms the ``quantile'' and ``least confidence'' strategies. In low data regimes finding any positive examples at all is a challenge and annotation efforts are best spent on the most likely candidates. But when positive examples are plentiful, the lower quantiles provide a mix of ambiguous examples and broad diversity, both of which help improve the classifier. As a result, the ``top 10 + quantile'' active learning strategy is able to draw from the strengths of both the ``most confidence'' and ``quantile'' strategies and achieves consistently strong performance across data regimes.

\begin{figure}[t]
	\centering
    \begin{subfigure}{0.4851\textwidth}
        \includegraphics[width=\textwidth]{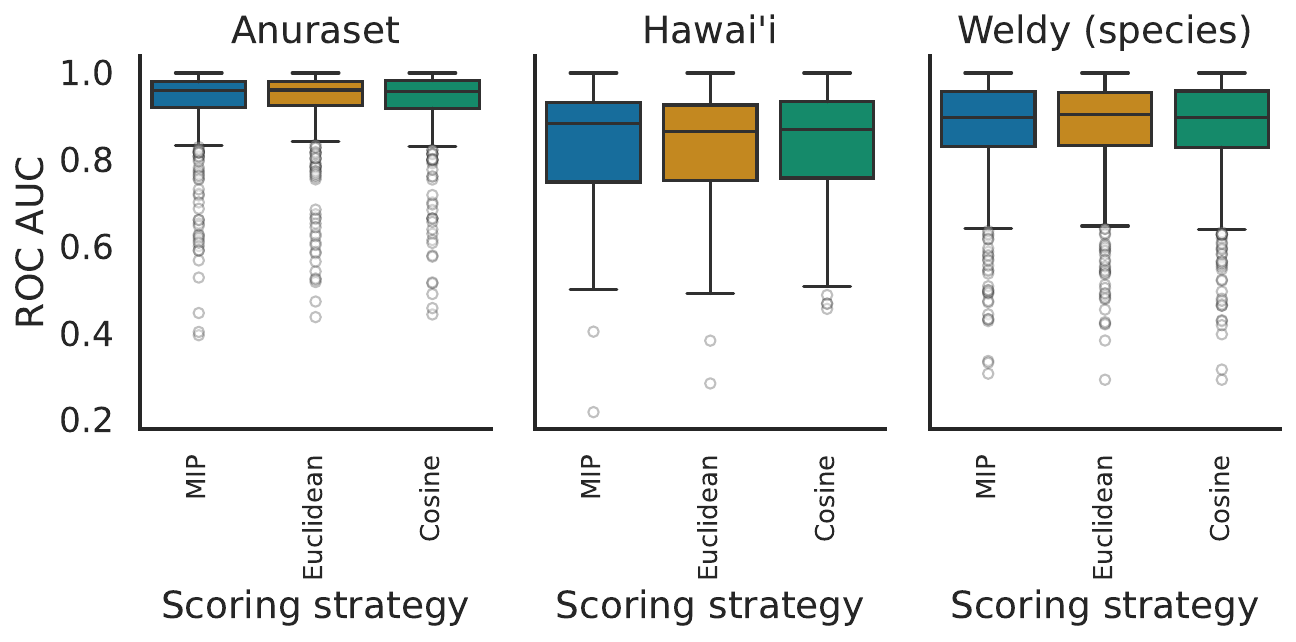}
        \caption{\label{fig:scoring_strategies} Embedding scoring in the initial search phase.}
    \end{subfigure}
    \hfill
    \begin{subfigure}{0.4782\textwidth}
        \includegraphics[width=\textwidth]{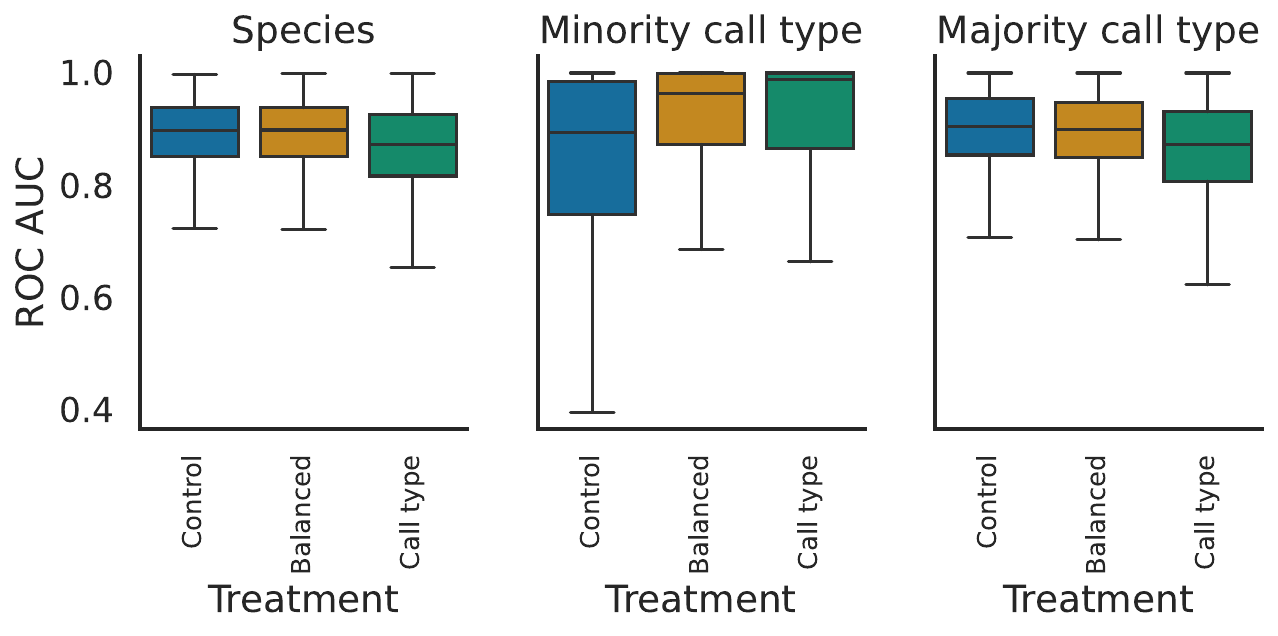}
        \caption{\label{fig:call_type_management_comparison} Call types handling.}
    \end{subfigure}
	\caption{Simulated agile modeling performance for Perch embeddings and a ``top 10 + quantile'' active learning strategy for various ways of (a) scoring embeddings in the initial search phase and (b) handling call types.}
\end{figure}

\subsubsection{Robustness of Initial Search Similarity Score Function}

Using the best-performing embedding function (Perch) and active learning strategy (top 10 + quantile), we examine how robust the initial search is to scoring strategies other than maximum inner-product, which we compare against Euclidean distance and cosine similarity across datasets and data regimes (\autoref{fig:scoring_strategies}). We find that the choice of scoring strategy does not significantly impact performance.

\subsubsection{Call type management: a balanced search query strikes a good compromise}

Again using the best-performing embedding function (Perch) and active learning strategy (top 10 + quantile), we finally compare the {\em Control}, {\em Balanced}, and {\em Call type} strategies for handling species with multiple call types (\autoref{fig:call_type_management_comparison}). Results vary from species to species, but at an aggregated level we observe that a call type-balanced search query followed by a species-level agile modeling loop ({\em Balanced}) generally provides better minority call type performance at a negligible majority call type and species-level performance cost. Conversely, splitting the annotation budget into independent call type-level agile modeling loops ({\em Call type}) provides an even greater minority call type performance improvement, but at the cost of majority call type performance.

\section{Discussion}

In this work, we have demonstrated accelerated development of classifiers of direct use for important questions in ecology and conservation. The rapid human evaluation of search results (4.79 seconds per clip) alongside rapid computational processes for search and classifier training allows direct estimation of the effort required to produce new classifiers: A review time of 4.79 seconds per clip indicates a human reviewer can process 720 examples per hour, sufficient to produce and evaluate a good-quality classifier in most cases.

\subsection{Honeycreeper Monitoring}

We have demonstrated that the agile modeling system can be used to quickly and efficiently assess the distribution and abundance of not only individuals of different species, but also individuals in different life stages. Our findings from Experiment B: Two Species Annotation align with abundance patterns observed in other studies that indicate ʻAkiapōlāʻau and ʻAlawī densities are increasing in high-elevation pasture areas, but are decreasing in lower-elevation closed forest areas~\cite{Kendall_2023}. Our study design does not allow us to disentangle elevation and ecotype effects on call densities, but still serves as a proof of concept for using call density to track how specific signals may vary across covariate gradients. Even at the most abundant sites call densities were relatively low for both species, despite restricting the dataset to only their peak pre-breeding vocalization windows. Phenology of vocalization windows for these species is poorly studied and understood, and may be shifting with climate or other ecosystem changes~\cite{Iler_2021}. This highlights another opportunity for the agile modeling system described here to fill critical knowledge gaps.

It was somewhat surprising that ʻAlawī call densities not only did not track those of the adults, but also did not appear to be correlated with elevation. It may be that ʻAlawī adults spend considerable time within the high-elevation koa pasture areas, but prefer to nest within mid-elevation open canopy forest. Alternatively, because juveniles stay close to their nest site during the fledging period, a time when their begging calls are particularly conspicuous~\cite{VanderWerf_1998,Ralph_Fancy_aki}, recorder placement near a nest may lead to higher call densities than are reflective of the study area. On the other hand, recorders within productive breeding habitat may yield deceptively low juvenile call densities if not placed close to any given nest. This effect may be greater for species with shorter periods of parental care, such as with the ʻAlawī~\cite{VanderWerf_1998}, but less so for species with longer juvenile dependency, like ʻAkiapōlāʻau, as the fledglings become more mobile and vocalize farther from their nests as they develop~\cite{Pratt_2020}. With this in mind, it may be better practice to assess across a study area with multiple recorders than to use a single recording site. 

For the purposes of this case study we did not include other honeycreeper species vocalizations in the training of the classifiers. During the validation process for adult ʻAlawī, annotators frequently observed Hawaiʻi ʻAmakihi (\textit{Chlorodrepanis virens}) vocalizations scoring in the top bins. Inclusion of acoustically similar species in the training data can improve confidence intervals around call density estimates by facilitating removal of false examples from the top bins, thereby providing better clustering of positive examples~\cite{Navine_2024}.

\subsection{Coral reef health}
Related work has relied on manual counts or single species classifiers to detect fish sounds~\cite{Raick_2023,Munger_2022}. The methodology presented in our study enabled the rapid development of highly accurate classifiers, with a minimum ROC-AUC of 0.98, for the nine fish sonotypes in 3.09 hrs of cumulative human labeling effort. The resultant classifiers revealed both a higher abundance and diversity of fish sonotypes on healthy sites compared to degraded sites. We also report comparable values for these two measures between the healthy baseline sites and actively restored sites. The increased abundance of detections was primarily driven by increases in the `Pulse train' and `Rattle' sonotypes. Whilst sonotypes  were not corroborated to specific taxa in the field, the former fits the description of pulse trains often associated with Damselfish (\textit{Pomacentridae})~\cite{Munger_2022}. Damselfish are a highly abundant and diverse group that deliver a host important ecological functions including algal grazing, territorial impacts and coral mutualism~\cite{Ceccarelli_2005,Pratchett_2012}. The rattle sonotype exhibited strong peaks in activity between the 7--11pm period for the days around the new moon period. This behavior is typical of spawning fish, which may indicate recovery of spawning by the responsible taxa on restored coral reefs~\cite{Mcwilliam_2017}. 

The Zero-Inflated Poisson (ZIP) analysis of the detection data shows clear differences between healthy and degraded sites for a subset of sonotypes. The ZIP distribution is itself a kind of occupancy model, with the $\pi$ parameter acting as a latent variable for presence of the vocalizing animal, even when detection rates are low.
Matching sonotypes to the vocalizing species can be extremely difficult in the reef context: Identifying indicator sonotypes allows better prioritization of investigative effort. These indicator sounds can demonstrate whether the restored reef sites are behaving similarly to healthy sites, and identifying the specific vocalizing species will provide further support for the ecological context of the detection metrics. For sonotypes where the restored sites behave more like the degraded sites (such as the `croak' call), identifying the vocalizing species will help understand aspects where the restored habitat is failing to provide replacement habitat.

Recent work has shown that bioacoustic signals are important parts of coral reef recruitment~\cite{mooney2024} and restoration. We expect that our approach will allow improved study of the impact of sound diversity and composition on coral reef recruitment.

\subsection{Christmas Island Species ID}

Despite the paucity of recordings in the training set for the embedding model, iterative active learning produced high-quality classifiers for all three target species (ROC-AUC greater than 0.95), with less than an hour of analyst time each. We also found that the system worked well with the large volumes of data involved. Working with uniformly random samples of the large dataset was sufficient to obtain high classification quality. The scarcest species was the Emerald Dove, with approximately 4\% call density, so any 10k hour uniformly random sample of data should contain around 400 hours of audio containing the Emerald Dove. However, for extremely scarce signals, an approximate nearest neighbor search which can work with the entire dataset may be preferable to a sampled approach.

The accuracy of the classifiers enabled us to extract data on vocal activity rates of multiple target species from a large set of daily recordings collected across Christmas Island over more than a year. This gives valuable insight into the spatial usage patterns of these threatened species which will inform monitoring decisions. The Christmas Island Emerald Dove and Christmas Island Goshawk, while widespread across the island, had very low detection rates at many sites. Selecting sites for monitoring based on highest detectability for these two species would also effectively capture other species, such as the Christmas Island Thrush, which have high detectability at most sites. In this case study we aggregated data across the entire 16-month monitoring period for a view into overall site usage. However, the high-quality output of the classifiers provides further opportunity to examine vocal activity patterns at finer temporal scales, including seasonal and diel patterns. The Christmas Island monitoring effort plans to use these models going forward to design a time- and cost-effective acoustic monitoring program that can detect declines or disappearance from sites. Understanding the variation in site activity for these species will help to prioritise locations where acoustic monitoring would be most effective for detecting changes in vocal activity over time.

\subsection{Practical recommendations}

Our simulated experiments reveal important practical insights for practitioners interested in taking advantage of the framework. Firstly, embedding function quality matters, and practitioners are better served by performing agile modeling on top of the more recent embedding functions trained on bird (and other animal) vocalization data, such as BirdNet v2.3, Perch, and SurfPerch. This is in agreement with transfer learning results presented in~\cite{ghani2023}, and consequently, as better embedding functions are made available their performance as measured by a (relatively simpler) transfer learning evaluation protocol should be indicative of their performance when incorporated into the agile modeling framework. Secondly, absent any prior information on abundance, a {\em top $k$ + quantile} active learning strategy is a robust and well-rounded way to decide which results to surface for human validation. Finally, incorporating expert knowledge into the search query in the form of a balanced set of query examples (in terms of vocalization types) is a simple but effective way of improving retrieval performance on minority vocalization types while maintaining strong performance on majority vocalization types.

\subsection{Conclusions}

In this work, we have demonstrated the broad efficacy of agile modeling for bioacoustics, building on foundational embeddings from bird classifiers. Our case studies demonstrate that this method meets the efficiency, adaptability, scalability, and quality criteria necessary to serve as a transformative ecological research and conservation tool. The extreme efficiency of model development for passive acoustic monitoring projects promises to allow the investigation of a much wider range of questions for which training data is scarce or non-existent, such as juvenile call monitoring for tracking population health and reproductive success, or monitoring extremely rare birds. We have shown that detection data from novel classifiers can be easily incorporated into the ecosystem understanding, as we saw for fish sounds in coral reefs. And with Christmas Island, we see that our system can scale easily to very large datasets. Moreover, our simulated experiments reveal important practical insights for practitioners interested in taking advantage of the framework.

At the same time, much remains to be done. In terms of scalability, the incorporation of approximate nearest neighbor (ANN) search will likely make agile modeling amenable to even larger applications. We conjecture the system would be robust to the imperfect recall intrinsic to ANN, however this remains to be verified empirically, and doing so systematically is challenging in and of itself. In terms of quality, the training of audio representations for bioacoustics remains an open research problem, and we expect future work in that direction to improve performance when incorporated into the system. In particular, we observe some variability of outcomes in our simulated experiments, and better embedding functions could help improve worst-case performance. More work is also needed on strategies to handle species with multiple vocalization types beyond our initial exploration. This is a question that often arises in practice, and we see great potential in exploring more sophisticated strategies. Finally, our case studies themselves raised interesting questions. For instance, the biological origin of the ``croak'' sound in the coral reefs study remains a mystery, and more field work is needed to determine what makes the sound in the reefs and why it is missing from restored reef sites.


We hope this work will improve our understanding of both terrestrial and marine ecosystems, and ultimately contribute to better management of endangered and invasive species.

\section{Acknowledgements}

Coral Reefs Case Study:
Logistical research support was provided by MSS. We also thank the Department of Marine Affairs and Fisheries of the Province of South Sulawesi, the Government Offices of the Kabupaten of Pangkep, Pulau Bontosua and Pulau Badi, and the community of Pulau Bontosua for their support. Fieldwork in Indonesia was conducted under an Indonesian national research permit issued by BRIN (number 108/SIP/IV/FR/2/2023, jointly held by T.A.C.L. as lead foreign researcher and T.B.R. as lead Indonesian host researcher), with ethical approval from BRIN and Lancaster University. We thank Prof. J. Jompa and Prof. R.A. Rappe at Universitas Hasanuddin for logistical assistance with permit and visa applications. Member of Mars Coral Restoration Project Monitoring Team: L. Damayanti, M. E. Prasetya, P. B. Maulana, A. Hamka, A. Dwiyanto, A. M. A. Pratama, A. T. Abeng, Irwan, R. Madjid, E. Agiel, C. V. Parrangan, H. Lakota, Hamzah, and Suandar. Funding was provided by a PhD studentship for the Fisheries Society of the British Isles to B.W and a research fellowship from The Royal Commission for the Exhibition of 1851 to T.A.CL.

Hawai`i Case Study: All audio used in this case study was transferred to a virtual computer hosted by Jetstream2~\cite{Hancock_2021} for analysis. We also thank Saxony Charlot, Braxton Igne, and Nikolai Braedt for significant contributions to annotation and data collection.

Christmas Island Case Study: Project management and collection of audio recordings was undertaken by Parks Australia Science Team with support from the Christmas Island National Park team. We thank Jess Williams, Lil Blake, Nick Macgregor, Margarita Goumas, and Brendan Tiernan at Parks Australia. We also thank Cecile Espigole and Ted Pedersen for assistance in the field. Field work was conducted under James Cook University Animal Ethics Approval number A2907. The project was supported with funding from the Australian Government under the National Environmental Science Program’s Resilient Landscapes Hub.

\printbibliography

\end{document}